\documentclass[10pt,conference]{IEEEtran}

\usepackage{graphicx}
\usepackage{epstopdf}
\usepackage{amsmath,amssymb,amsthm}
\usepackage{cite}
\usepackage{float}
\usepackage{comment,enumerate}
\usepackage{color}
\usepackage{algpseudocode}
\usepackage{algorithm}
\usepackage[utf8]{inputenc} 
\usepackage[T1]{fontenc}
\usepackage{mathtools}
\usepackage{courier}
\usepackage{array} 
\newcolumntype{L}{>{$}l<{$}} 

\newcommand{\ieb}{\begin{IEEEeqnarray}{rCl}}
	\newcommand{\ien}{\end{IEEEeqnarray}}

\usepackage{comment}
\title{Joint Scheduling and Throughput Maximization in Self-backhauled Millimeter Wave Cellular Networks }
\author{\IEEEauthorblockN{Chao Fang\IEEEauthorrefmark{1}, Charitha Madapatha\IEEEauthorrefmark{1},
		Behrooz Makki\IEEEauthorrefmark{2}, and
		Tommy Svensson\IEEEauthorrefmark{1}}	
	    \IEEEauthorblockA{\IEEEauthorrefmark{1} Department of Electrical Engineering,
		Chalmers University of Technology, Gothenburg, Sweden\\
		\IEEEauthorrefmark{2} Ericsson Research, Ericsson AB, Gothenburg, Sweden\\
		Email: \IEEEauthorrefmark{1} \{fchao, charitha, tommy.svensson\}@chalmers.se, \IEEEauthorrefmark{2} \{behrooz.makki\}@ericsson.com
	}
}

\begin{document}
	\maketitle
\begin{abstract}
	Integrated access and backhaul (IAB) networks have the potential to provide high data rate in both access and backhaul networks by sharing the same spectrum. Due to the dense deployment of small base stations (SBSs), IAB networks connect SBSs to the core network in a wireless manner without the deployment of high-cost optical fiber. As large spectrum is available in mmWave bands and high data rate is achieved by using directional beamforming, the access and backhaul links can be integrated in the same frequency band while satisfying quality-of-service constraints. In this work, we optimize the scheduling of access and backhaul links such that the minimum throughput of the access links is maximized based on the revised simplex method. By considering a probability based line-of-sight (LOS) and non-line-of-sight (NLOS) path loss model and the antenna array gains, we compare the achievable minimum access throughput of the IAB network with the network with only macro base stations, and study the effect of the network topology and antenna parameters on the achievable minimum throughput. Simulation results show that, for a broad range of parameter settings, the implementation of IABs improves the access minimum achievable throughput.
	
\end{abstract}
\section{Introduction}
	The evolution of wireless communication technology has led to advances in many areas.
	Today, 5G has become a key enabler for applications such as virtual reality, vehicle-to-everything communications, massive machine-type communications, etc. The growth in the demand of data rate, reliability and the number of connections require 5G to support high-speed, low-error-rate and low-latency communications with massive connectivity. Compared to 4G Long-term-evolution (LTE), 5G development includes many novel designs, in terms of radio architecture, carrier frequency and network deployment, among which the three most important technologies are millimeter-wave (mmWave) communications, massive multiple-input and multiple-output (MIMO) and network densification \cite{J_5G,Shafi_5G,rajatheva2021scoring}. 
	
	The large amount of spectrum available at mmWave frequency bands makes it possible to achieve a data rate as high as giga-bit-per-second. However, due to the high attenuation of mmWave signals and sensitivity to blockage, the range of mmWave communication is limited to a few hundred meters and the signal strength varies greatly depending on the line-of-sight (LOS) or non-line-of-sight (NLOS) conditions \cite{Bai_CoverageandRate, willwork}. The drawbacks of mmWave communications are expected to be reduced through beamforming with massive MIMO antenna arrays and network densification. Due to the small wavelength of mmWave signals, large antenna arrays have a small footprint, which makes beamforming well suited for high-capacity transmission \cite{Adhikary_mmWaveSDM, Sohrabi_HybridLargeScaleAntenna}. Also, high density deployments of small cells can greatly improve the NLOS situations and reduce the path loss \cite{Bai_CoverageandRate, singh_selfbackhaul, XinchenZhang_multislopePL,fang2021hybrid}.
	
	With the dense deployments of small cells, a key challenge is to provide a large amount of backhaul data into the core network. Traditional solutions with fiber connections between small cells and a macro cell may become infeasible due to the high cost, long installation time and low flexibility.
	 With this background and also motivated by the fact that the backhauling to small cells needs to support standardazied  NLOS connections, integrated access and backhaul
	(IAB) networking has recently received considerable attention \cite{ge_backhaul, Behrooz_IAB,madapatha2021topology, 9187867}. With IAB, the objective is to provide flexible wireless
	backhauling using 3GPP new radio (NR) technology in international mobile telecommunications (IMT) bands, and provide not
	only backhaul, but also the existing cellular services in the
	same node and via the same hardware. Such a network not only increases the flexibility but also may reduce the implementation cost and time-to-market.

	In this paper, we study the minimum achievable access throughput of multi-hop IAB networks. Considering a probability based LOS and NLOS path loss
	model, we utilize the revised simplex method to derive a proper scheduling and resource allocation scheme maximizing the minimum access rate. Moreover, we evaluate the effect of different parameters such as blockage, antenna gain and network density on the system performance, and compare the results with those achieved in the cases with only macro base stations (MBSs). The simulation results show that, despite of the additional resources needed for backahuling, IAB networks with the optimal scheduling and resource allocation can provide a higher
	achievable minimum throughput than that of the macro-only
	networks for a broad range of MBS transmit powers, especially in dense blockage environments. Also, we show that the achievable minimum throughput increases with network desification and refined antenna parameters.

\section{System model}
	
\begin{figure}[t]
		\centering
		\includegraphics[width=1.1\columnwidth]{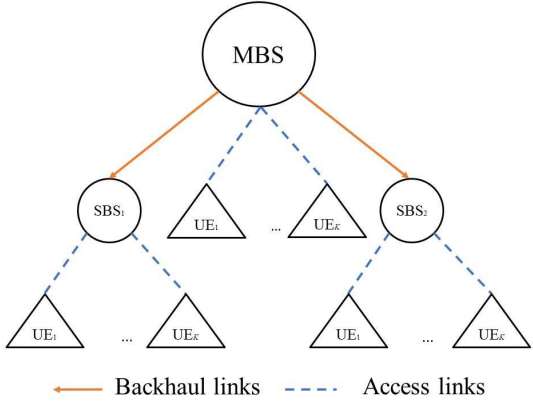}
		\caption{A wireless backhaul and access network example.}
		\label{network1}
	\end{figure}

	We consider a multi-hop downlink mmWave IAB network with $M$ MBSs and $R$ small base stations (SBSs). The BSs serve a total of $K$ users in the access network. The MBSs are connected to the core networks via non-IAB backhauling, e.g., fiber, while the SBSs are connected to the MBSs and receive backhauling data wirelessly.
	We use graphs to represent the network topology where the BSs or users that receive data from BS $m$ are treated as the child nodes of BS $m$, and the maximum number of child nodes per BS is denoted by $C$. The network expands from the MBSs in a tree structure such that there are $R+K$ links in the network. An example network with $M=1, R=2, C=2$ is shown in Fig.~\ref{network1}. The SBS nodes receive backhauling data from the MBS and each SBS maintains a number of a ccess links in order to serve the user nodes. 
	
		The BSs and users are assumed to be equipped with antenna arrays and use directional beamforming for array gain. We assume that the array gain is $M_t$ for all angles in the main lobe of the transmission antenna arrays and $m_t$ otherwise. Similarly, the $M_r$ and $m_r$ are the antenna gain for receivers. The beamwidth of the main lobe in the transmitter and receiver are denoted as $\theta_t$ and $\theta_r$, respectively. We assume that for the desired link, the maximum array gain is always achieved while the array gain is a random variable $D_{j,k}$ for any interfering link between nodes $j$ and $k$. We assume that angles of departure (AOD) and angles of arrival (AOA) are uniformly distributed in $(0,2\pi]$. Hence, the array gain of interfering links is decided depending on if the AOD and AOA falls within the main lobe of the transmitter and receiver.
	
		Each link can be LOS or NLOS according to the blockage situations, we use a random LOS model to calculate the path loss of the links. The path loss between node $m$ and $k$ is given by
	\ieb
	l_{m,k}=\mathbb{I}(p_L(d))\mathrm{PL}^{-1}_{\textrm{LOS}}
	+\left(1-\mathbb{I}(p_L(d)) \right)\mathrm{PL}^{-1}_{\textrm{NLOS}},
	\label{pathloss}
	\ien
	where $d$ is the distance between node $m$ and $k$, $\mathbb{I}(p_L(d))$ is a Bernoulli random variable with LOS probability $p_L(d)=e^{-\beta d}$, and $\beta$ models the blockage density \cite{Bai_CoverageandRate}. $\mathrm{PL}_{\textrm{LOS}}$ and $\mathrm{PL}_{\textrm{NLOS}}$ are the path loss for LOS and NLOS links, respectively. We use the close-in free space reference distance model given by 
	\ieb
	\mathrm{PL}[\mathrm{dB}] (d)= 20\log_{10}\left(\frac{4\pi}{\lambda}\right) + 10 \alpha_{\mathrm{LOS}/\mathrm{NLOS}} \log_{10}(d) \nonumber \\
	+ X_{\sigma_{\mathrm{LOS}/\mathrm{NLOS}}}, d\ge 1~\mathrm{m},	
	\ien
	where $\alpha_{\mathrm{LOS}/\mathrm{NLOS}}$ is the path loss exponent for LOS/NLOS links and $X_{\sigma_{\mathrm{LOS}/\mathrm{NLOS}}}$ is the log-normal distributed shadowing parameter with standard deviation $\sigma_{\mathrm{LOS}/\mathrm{NLOS}}$. In addition to the path loss, we assume independent Nakagami fading on each link and the signal-to-interference-plus-noise ratio (SINR) for a link between nodes $m$ and $k$ is given by
	\ieb
	\textrm{SINR}_{m,k}= \frac{M_rM_tg_{m,k}l_{m,k}}{\sigma^2+ \sum_{j,j\ne m}D_{j,k}g_{j,k}l_{j,k}},
	\ien 
	where the Nakagami fading variable $g_{j,k}$ follows the gamma distribution $p_g(x) = q^q \frac{x^{(q-1)}}{\Gamma(q)}\exp(-qx)$ with factor $q = 1/N_{\text{LOS/NLOS} }$, $N_{\text{LOS}}$ and $N_{\text{NLOS}}$ model the fluctuations of the signal strength due to  multipath scattering for LOS and NLOS links, respectively,
	and $\sigma^2$ is the noise power. Therefore, assuming Gaussian signaling and for unit bandwidth, the link capacity is given by $c_{m,k}=\log_{2}(1+\textrm{SINR}_{m,k})$ between link $m$ and $k$. The users are assumed to be associated to the BS with maximum received power averaged over fading. Therefore, BS $m$ serves user $k$ if $l_{m,k} = \max_{i=1}^{M+R}l_{i,k}$.
	
	We assume that the BSs can only transmit or receive, i.e., not simultaneously, in a given time slot and the BSs transmit to their serving users and BSs in a time-division multiple access (TDMA) manner. Hence, only one of the links connected to a BS can be active and the active links can not share a common BS at a given time.
	In order to find the optimal scheduling that maximizes the user throughput, we divide each frame into a number of time slots with variable length. In each time slot, a certain set of links are active. Therefore, the optimal scheduling problem is equivalent to finding the optimal time resource allocation to every possible set of activation links.
	We denote the duration of a time slot as $t_i$. By defining a unit time frame, we have $\sum_{i=1}^{N}t_i = 1$ where $N$ is the total number of time slots in a unit frame.

\section{Minimum throughput maximization}	
	
	 In this section, we formulate the optimization problem which maximizes the minimum throughput of each access link subject to the time resource constraint. Such an optimization guarantees better fairness in the network, at the cost of losing the peak access rate for the users with good link quality.

	 Let $S$ be a schedule where in each time slot a certain set of links are active. The throughput of a link between nodes $m$ and $k$ is given by $\gamma_{m,k} = \sum_{i=1}^{N} t_i c_{m,k}$. If the link $m\rightarrow k$ is not active in time slot $t_i$, we assume $ c_{m,k}=0$. Hence, the total throughput delivered to a node $k$ is given by 
	 \ieb
	 	r_k &=& \sum_{m\in B_k} \gamma_{m,k} - \sum_{n\in A_k}\gamma_{k,n}, \\
	 	    &=&  \sum_{i=1}^{N} t_i\left ( \sum_{m\in B_k}c_{m,k} - \sum_{n\in A_k} c_{k,n} \right)
	 \ien
	 where $B_k$ is the set of IAB nodes that provide backhaul data to node (either in access or backhaul) $k$ and $A_k$ is the set of nodes to which node $k$ transmits data. Denoting $\mathbf{t} = [t_1,\cdots,t_N]$ and $\mathbf{C}=[\mathbf{c}_1,\cdots,\mathbf{c}_i, \cdots,\mathbf{c}_N]$ as the capacity matrix where the elements in $\mathbf{c}_i$ are the aggregated rates of each node based on a given link activation pattern in time slot $t_i$, i.e., $\mathbf{c}_i[k] = \sum_{m\in B_k}c_{m,k} - \sum_{n\in A_k} c_{k,n}, \forall k$. 
	 
	 In general, assume node $k$ is required to achieve a minimum throughput $\theta_k$, our objective is to maximize the minimum throughput of the nodes in the access network subject to the time resource constraint. The problem can be formulated as	
	\ieb
	\label{objective}
	&&\mathcal{P}:\mathrm{max}~~~\theta\\
	&&\mathrm{s.t.}~~~ \mathbf{C}\mathbf{t} \ge \theta \mathbf{w}_{\theta}\\
	&&\mathbf{1}^T\mathbf{t}=1\\
	&&\mathbf{t} \ge \mathbf{0},
	\ien 
	where $\theta \mathbf{w}_{\theta}$ is the minimum throughput of all nodes and $\mathbf{w}_{\theta}[k]$ is the weight for node $k$ in order to achieve different minimum throughput among nodes and provide larger throughput for users with higher data demand. 
	Here we are interested in maximizing the minimum throughput to the users. To that end, for BS nodes we can simply set the weights to be 0 so that the BSs are treated as gateways to provide throughput to the users. 
	
	Problem $\mathcal{P}$ is difficult to solve, because, in each time slot $t_i$, there exists a number of feasible link activation patterns, and thus $\mathbf{c}_k$ varies with different link activation patterns. 
	In order to find the optimal schedule in which $\mathbf{C}\mathbf{t}$ gives the maximum throughput of each users, we propose an iterative algorithm based on the revised simplex algorithm\cite{Yuan_jointrouting} and maximum matching theory \cite{mwm, rsm}. We first create an initial feasible schedule and $\mathbf{C}$, then iteratively update the schedule and $\mathbf{C}$ until there is little or no improvement to the throughput. 
	
	First, we convert the linear program $\mathcal{P}$ to its canonical form by introducing surplus variables. The equivalent problem can be written by
	\ieb
		&&\mathcal{P}_0:\mathrm{min}~~~\mathbf{f}^T\mathbf{x}\\
		&&\mathrm{s.t.}~~~ \mathbf{U}\mathbf{x} =  \mathbf{g}\\
		&&\mathbf{x} \ge \mathbf{0},
	\ien
	where 
	\ieb
		\mathbf{U} = 
					\begin{bmatrix}
						\mathbf{C}& \mathbf{- \mathbf{w}_{\theta}} & -\mathbf{I} \\
						\mathbf{1}^T& 0           & \mathbf{0}^T						
					\end{bmatrix},
	\ien
	 $\mathbf{f}^T=\begin{bmatrix} \mathbf{0}^T & -1 &\mathbf{0}^T \end{bmatrix}, \mathbf{x}^T=\begin{bmatrix} \mathbf{t}^T & \theta &\mathbf{s}^T \end{bmatrix}, \mathbf{g}^T=\begin{bmatrix} \mathbf{0}^T & 1 \end{bmatrix}$ and $\mathbf{s}$ are the surplus variables.
	 
	In order to solve $\mathcal{P}_0$, conditioning on $N=R+K$, we first find an initial schedule in which only one of the links is active in each time slot. Since the network consists of $R+K$ links, in each time slot $t_i$, a different link is active and $\mathbf{c}_i, \forall i$ is computed according to $\mathbf{c}_i[k] = \sum_{m\in B_k}c_{m,k} - \sum_{n\in A_k} c_{k,n}, \forall k$.
	Based on the initial $\mathbf{C}$, $\mathcal{P}_0$ can be solved via the simplex method and we obtain the optimal basis $B$. Next, we iteratively update the schedule and $\mathcal{C}$ based on revised simplex method and the maximum matching theory \cite{Yuan_jointrouting}. 
	
	We then update $\mathbf{C}$ according to
	\ieb
	\label{weight}
		c_{m,k}= \begin{cases}
				c_{m,k}  (p_m-p_k)~~ \text{if}~\text{nodes $m,k$ are SBSs or users}~ \\
				c_{m,k}  p_m ~~~~\text{otherwise,}
			
		\end{cases}
	\ien
	where $p_m$ is the $m$-th element in $\mathbf{p}=\mathbf{f}_B^T B^{-1} $ and $\mathbf{f}_B$ are the coefficients in $\mathbf{f}^T$ corresponding to the initial basis. According to maximum matching with $c_{m,k}$ being the weights of each link, we obtain a new schedule where in each time slot, an updated set of links are active. We then compute 
	\ieb
		&&\eta_1 = -\sum_{m} c_{m,k} -p_{R+K+1},~~~\text{for all MBS nodes $m$ } \\
		&&\eta_2 = -1+ \sum^{R+K}_{i=1} p_i\\
		&&\eta_3 = \min_{1\le i \le R+K} p_i\\
		&&\eta= \min(\eta_1,\eta_2,\eta_3).
	\ien
	If $\eta\ge 0$, then the global optimal  $\theta$  is found, otherwise $B$ is updated by solving $\mathcal{P_0}$ with updated $\mathbf{C}$. With each iteration, we compute a new $\mathbf{C}$ with a new schedule and the throughput are improved, the stopping criterion applies if there is no further improvement of the throughput. Problem $\mathcal{P}_0$ are guaranteed to converge in polynomial time \cite{converge}.

	It is worth noting that the algorithm assume that there is always enough data at each BS to transmit to other SBSs or users. This may not be true in the initial few time slots, however, by assuming a data buffer in each BS, the maximum throughput is achieved on average.
	\section{Simulation Results}
	
	In this section, we present the simulation results based on the proposed minimum throughput optimization algorithm and show that the achievable minimum throughput of IAB networks with the optimal scheduling and resource allocation can outperform that of macro-only networks for a broad range of network parameters. In addition, we analyze the effect of the network, blockage and the array gain on the achievable minimum throughput.
	
	\begin{figure}[t]
		\centering
		\includegraphics[width=\columnwidth]{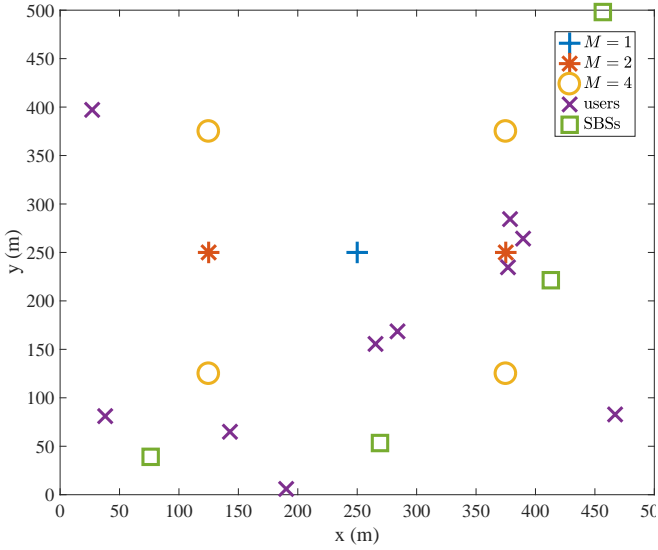}
		\caption{An example realization of the locations of the BSs and users. `$M=1,2,4$' denotes the locations of the MBSs when the number of MBSs is $1,2,4$, respectively. Here, $K=10$ and $R=4$.}
		\label{bsloc}
	\end{figure}  
	
	The locations of the MBSs are predefined and the locations of SBSs and users are randomly generated in a 2D plane, one example realization of the network is shown in Fig.~\ref{bsloc}. We assume the maximum node degree $C=2$, i.e., each BS can provide backhauling to a maximum of 2 BSs, and the backhaul network expands from the MBSs. The mmWave network is assumed to operate at 28 GHz with bandwidth 100 MHz per link. The transmit power of SBSs is $P_{\mathrm{SBS}}= 30$ dBm and the transmit power of MBSs is $P_{\mathrm{MBS}}= 40$ dBm unless otherwise stated. The LOS and NLOS path loss exponents and the LOS and NLOS log-normal shadowing standard deviations are $\sigma_{\mathrm{LOS}} = 3.6$ dB, $\sigma_{\mathrm{NLOS} }= 9.7$ dB, respectively. The Nakagami fading scaling parameters are $N_{\text{LOS}} =3$ and $N_{\text{NLOS}} = 2$. The LOS probability parameter is $\beta = 0.01$ unless otherwise stated. The backhaul links are assumed LOS while the access links and the interfering links follow the random LOS/NLOS path loss model. 
	The array gains are $M_t=M_r= 10$ dB and $m_t=m_r = -10$ dB unless otherwise stated and the main lobe beamwidths are $\theta_t= 30^{\circ}$ and $\theta_r=90^{\circ}$. The simulation results are averaged over $10^5$ channel realizations.

			 \begin{figure}[t]
		\centering
		\includegraphics[width=\columnwidth]{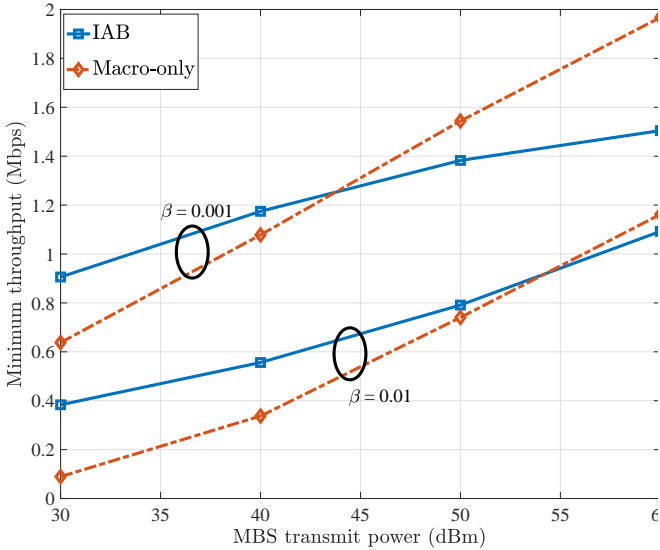}
		\caption{Minimum throughput versus the MBS transmit power. The network parameters for the IAB network are $M=1$, $R=2$, $K=10$, and $P_{\mathrm{SBS}}= 30$ dBm. The macro-only network consists of only $M=1$ MBS serving $K=10$ users. $\beta$ controls the LOS probability with a larger $\beta$ defining a denser blockage environment.}
		\label{noniab}
	\end{figure} 
	
	\begin{figure}[t]
		\centering
		\includegraphics[width=\columnwidth]{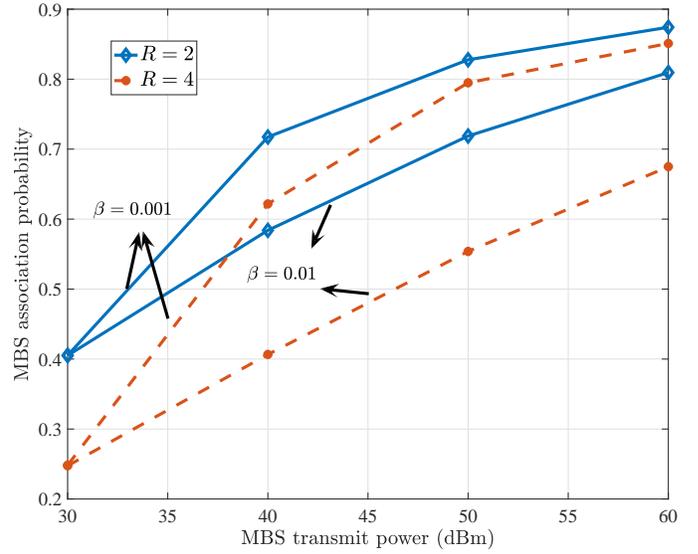}
		\caption{MBS association probability versus the MBS transmit power. The network parameters for the IAB network are $M=1$, $K=10$, and $P_{\mathrm{SBS}}= 30$ dBm. $\beta$ controls the LOS probability with a larger $\beta$ defining a denser blockage environment.}
		\label{usermum}
	\end{figure}

	 Figure~\ref{noniab} compares the achievable minimum throughput between IAB and macro-only networks for different MBS transmit power levels. It can be seen that the achievable minimum throughput for both IAB and macro-only networks increases with the macro BS power level, as a result of increased desired link power. 
	 Although IAB networks require additional resources on the backhaul links, the IAB network has a higher achievable minimum throughput than that of the macro-only network for a broad range of MBS transmit powers, especially in denser blockage environments. This is due to the reduced path loss and increased LOS probability for access links. The achievable minimum throughput of the macro-only network outperforms that of the IAB networks only when the MBS power level is considerably larger than the SBS power level (approximately 20 times for $\beta=0.001$ and 250 times for $\beta=0.01$), since larger MBS power causes more interference to users associated with SBSs in IAB networks.
	  
	 In Fig. \ref{usermum}, we show the probability that a user is associated to the MBS with different MBS power levels and SBS density. As expected, the MBS association probability increases with the MBS transmit power, LOS probability and decreases with the number of SBSs, as it is more likely for a user to be associated to BSs with shorter link distance and less path loss.
	 \begin{figure}[t]
	 	\centering
	 	\includegraphics[width=\columnwidth]{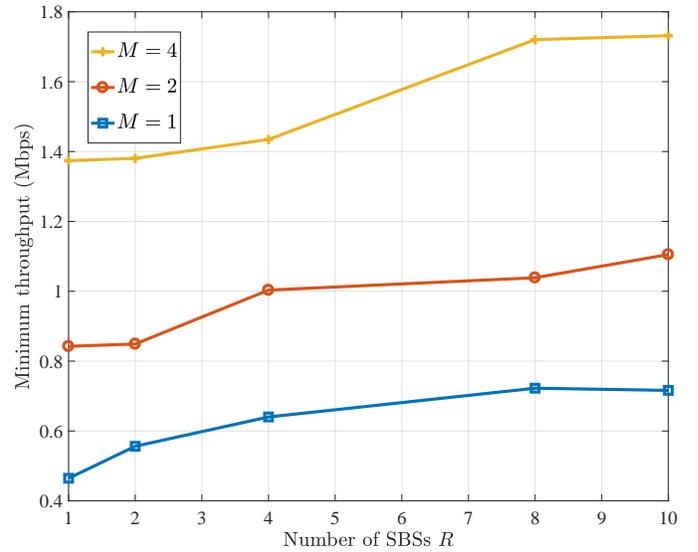}
	 	\caption{Minimum throughput versus the number of SBSs. The number of users is $K=10$.}
	 	\label{power}
	 \end{figure}  
 
	 In Fig~\ref{power}, we show the achievable minimum throughput for different numbers of MBSs and SBSs. The achievable minimum throughput increases with the number of SBSs $R$, as the network is further densified with SBSs and the links have shorter link distance, lower path loss and higher probability for LOS conditions on average. Also, we observe that the achievable minimum throughput increases with the number of MBSs $M$. This is intuitive because the number of backhauling hops decreases with $M$, since the SBSs are partitioned into IAB sub-networks starting from the MBSs. Therefore, less time resources are spent in backhauling links. It is worthwhile to mention that as the total number of BSs approach or exceeds the number of users, there is a decrease in the incremental throughput. This is because, according to the maximum received power association rule, some BSs have no attached users and serve as relay nodes or become inactive, when the number of SBSs increases.

	  \begin{figure}[t]
	 	\centering
	 	\includegraphics[width=\columnwidth]{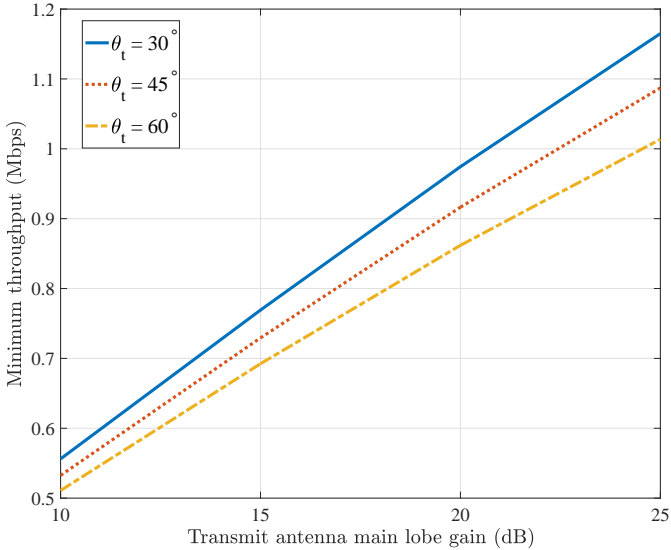}
	 	\caption{Minimum throughput versus the main lobe array gain. The number of users is $K=10$, the number of MBSs is $M=1$, and the number of SBSs is $R=2$.}
	 	\label{mt}
	 \end{figure}
 
 	The achievable minimum throughput for different transmit antenna main lobe gain and beamwidth is shown in Fig.~ \ref{mt}. Here, the results are presented for the cases with the side lobe gains of the transmitters and the antenna parameters of the receivers are fixed. It can be seen that larger minimum throughput can be achieved by increasing the main lobe gain or decreasing the main lobe beamwidth. Particularly, for the considered parameter settings, the minimum access rate increases almost linearly with the transmit antenna main lobe gain. Also, the relative throughput gain when increasing the main lobe gain increases as the main lobe beamwidth decreases.
 	
 	\section{ Acknowledgment}
 	This work was supported in part by VINNOVA (Swedish Government Agency for Innovation Systems) within the VINN Excellence Center ChaseOn.
	  
	 \section{Conclusion}
	 In the paper, we proposed a minimum throughput maximization algorithm for mmWave IAB networks. Our model considered the path loss difference due to blockages and the antenna array gains depending on if AODs or AOAs are within the main lobe. Based on the considered channel model, we maximized the minimum throughput of each access link based on the revised simplex method. The simulation results show that the achievable minimum throughput of IAB networks outperforms that of the macro-only networks for a broad range of MBSs power levels and IAB networks can provide better throughput in denser blockage environments. Moreover, we showed that the achievable minimum throughput increases when densifying the network with SBSs, reducing the number of backhaul hops, increasing the transmit antenna main lobe gain or reducing the main lobe beamwidth.
	 
	\vspace{-0mm}
	\bibliographystyle{IEEEtran} 
	\bibliography{ISWCS}
	\vfill
\end{document}